# Coalescence of Carbon Atoms on Cu (111) Surface: Emergence of a Stable Bridging-Metal Structure Motif

Ping Wu, Wenhua Zhang, Zhenyu Li,* Jinlong Yang, and Jian Guo Hou

*Hefei National Laboratory for Physical Sciences at Microscale, University of Science and Technology of China, Hefei, Anhui 230026, China*

By combining first principles transition state location and molecular dynamics simulation, we unambiguously identify a carbon atom approaching induced bridging metal structure formation on Cu (111) surface, which strongly modify the carbon atom coalescence dynamics. The emergence of this new structural motif turns out to be a result of the subtle balance between Cu-C and Cu-Cu interactions. Based on this picture, a simple theoretical model is proposed, which describes a variety of surface chemistries very well.

Graphene has attracted an intense research interest recently,[1] while massive production of high quality graphene sample remains to be a big challenge. Chemical vapor deposition (CVD) is a promising way to prepare graphene at large scale.[2] Among the studied metals, copper was recently recognized as the best substrate material for CVD growth of graphene.[3] At the same time, CVD is also widely used to synthesize carbon nanotubes (CNTs), where Cu can also act as the catalyst.[4] Since their solubility in Cu is very small, carbon atoms remain on the surface during the growth of graphitic materials.[5] It is thus essential to understand the atomic surface processes.

Recent studies on this topic mainly focus on the adsorption behaviours of different carbon species (mainly monomer and dimmer) at different Cu surface sites.[6,7] The carbon coalescence processes to produce larger species were traditionally expected to be straightforward without any covalent bond breaking.[8] However, this is not necessarily true, since the adsorbate induced geometric effects can be very significant.[9]

Considering the negligible diffusion barrier of a carbon atom on Cu surface,[6] a trivial carbon incorporation process will make dehydrogenation of methane the only difficult step during the growth of graphene. Kinetics of graphene growth is expected to strongly depend on the dynamics of carbon incorporation, and it is thus very important to unveil the previously overlooked atomic details of this process. In this Letter, we perform a direct carbon atom combination dynamics study via transition state location and molecular dynamics (MD) simulation. A new structural motif is unambiguously identified, which strongly modify the coalescence dynamics of carbon atoms on Cu (111) surface.

DFT calculations were performed with the Vienna *ab initio* simulation package (VASP)[10] using the PW91 exchange-correlation functional.[11] A four-layer slab model was used to describe the Cu (111) surface, in which the bottom layer was fixed to the optimized bulk geometry. The repeated slabs were separated by more than 10 Å to avoid interactions between neighbouring slabs. Plane wave cutoff and *k*-point sampling were carefully chosen to produce converged results. The climbing image nudged elastic band (CI-NEB) method[12] was used for transition state search and barrier height determination. Molecular dynamics (MD) simulations[13] were performed in the canonical

(NVT) ensemble using a time step of 1.0 fs. Scanning tunneling microscopy (STM) images were simulated using the Tersoff-Hamann model.[14]

We first put two carbon atoms on two neighbouring hollow sites. As previously assumed, they automatically form a carbon dimer without any barrier. At the same time, the calculated diffusion barrier for an individual surface carbon atom on Cu (111) surface is as small as 0.06 eV, agreeing well with previous results.[6] Therefore, it seems that we have now reached a conclusion supporting the previous assumption of trivial carbon incorporation.[8] However, there is an important point missed here. Before two carbon atoms reach neighbouring sites, they should pass though two next nearest neighbouring (*nnn*) sites (Figure 1a). Since two *nnn* carbon atoms already share a Cu atom, their interaction with the substrate may be different from that of individual adatoms.

Therefore, we also perform a geometry optimization for two *nnn* carbon atoms. A totally new structure (Figure 1b) is then obtained upon structure relaxation, where the surface Cu atom shared by the two carbon atoms is pulled upward from the surface. At the same time, one of the two carbon atoms (carbon A) penetrates into the subsurface through the hole left behind this Cu atom, forming an almost linear C-Cu-C configuration. Accompanied with the increases of carbon coordination numbers, the adsorption energy of this new bridging-metal (BM) structure becomes 1.01 eV larger than that of two isolated carbon atoms.

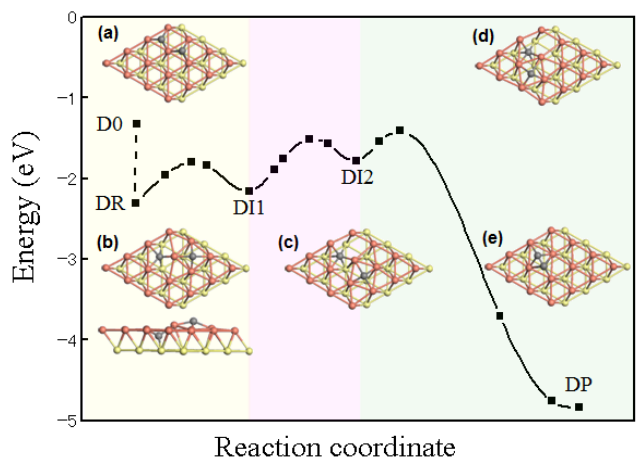

*Figure 1.* Minimum energy path of the 1+1 reaction. Optimization starting from D0 (a) automatically goes to the reactant DR (b). There are two intermediates, DI1 (c) and DI2 (d), between DR and the product DP (e).

With this BM structure, combination of the two carbon atoms (the 1+1 reaction) is stopped by the bridging Cu atom. It can proceed only if one of the carbon atoms makes a detour. After checking several possibilities, we have identified the minimum energy path (MEP) of this reaction, which is composed of three steps. First, the carbon atom remaining on the surface (carbon B) rotates around the bridging Cu to its neighbouring site (Figure 1c), with a 0.51 eV barrier. Then, it rotates further towards carbon A with an activation energy of 0.64 eV. Finally, by conquering a 0.37 eV barrier, carbon B drags carbon A to the surface and forms a dimer (Figure 1e). This MEP with several barriers thus gives a very rugged part of the two-adatom potential energy surface, which has never been noticed before.

To further study the early stage of graphene growth, we check how a third carbon atom can be attached to a preformed carbon dimer (the 2+1 reaction). We find that the diffusion of a dimer on the surface has only a moderate energy barrier (0.44 eV). Therefore, both dimer and atom can diffuse on Cu surface at a relatively high temperature.

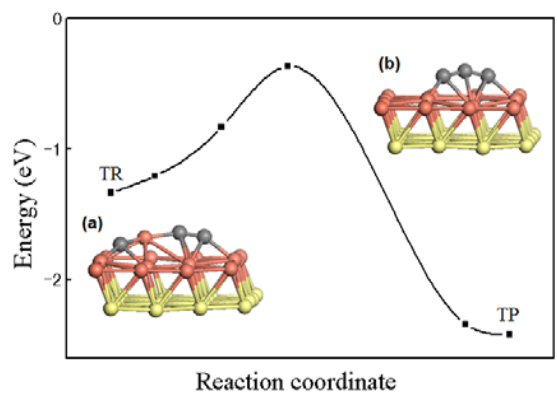

**Figure 2.** Minimum energy path of the 2+1 reaction. The reactant is TR (a), and the product is TP (b).

When the carbon atom approaches the dimer, a similar BM structure automatically forms, with the bridging Cu notably upshifted (Figure 2a). In contrast with the two-monomer case, no carbon atom penetrates into the subsurface. However, as an important feature of the BM structure, coordination number of the surface carbon atom has increased to 4. This BM structure gains a 0.41 eV additional adsorption energy compared with individual dimer and monomer. Product of the 2+1 reaction is a linear trimer (Figure 2b), the only stable isomer on Cu surface. Due to the formation of the BM structure, barrier of the 2+1 reaction (0.96 eV) becomes much larger than the diffusion barrier of either dimer or monomer. There is another approaching configuration for the 2+1 reaction, which gives similar results.

Four carbon atoms can form a triangle structure, with a central carbon atom occupying a top site on the surface. This structure is interesting since it is the smallest unit with a $sp^2$ carbon. The corresponding 3+1 reaction to form such a structure still has a BM-structured reactant (Figures 3a and 3b), which leads to a 1.24 eV barrier. Additionally, when a carbon atom approaches a preformed carbon hexagon, the BM structure is also observed (Figures 3c and 3d). Therefore, it is now safe to conclude that carbon incorporation is generally not barrierless on Cu surface, and it should be directly considered in studies of graphene growth dynamics.

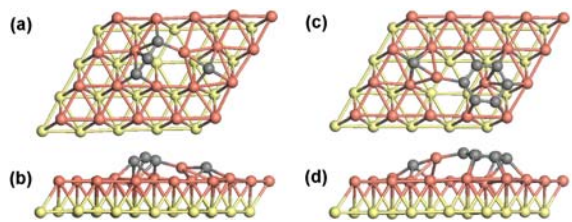

**Figure 3.** (a) Top view and (b) side view of the reactant of the 3+1 reaction, which is also with a BM structure. (c) Top view and (c) side view of the 6+1 BM structure.

Then, why approaching of carbon atoms automatically leads to the formation of BM structures? We propose the following mechanism. When two carbon atoms occupy two *nnn* sites, the shared Cu atom gets a large coordination number and becomes relatively unstable. If Cu-C interaction is preferred to Cu-Cu interaction, this Cu atom will be upshifted to make stronger Cu-C bonds, which generally also leads to increases of the coordination numbers of carbon atoms. That is to say, competition between Cu-C and Cu-Cu interactions may be the driving force behind BM structure formation.

To confirm this mechanism, we have checked several other metal (M) surfaces, as listed in Table 1. BM structures are found to be automatically formed on Cu, Ag, and Ni surfaces. As an indictor of the relative strength of M-C and M-M interactions, we calculate the ratio ($R$) between the carbon atom adsorption energy and the cohesive energy of the metals. Cu, Ag, and Ni have a relatively large $R$ value (>1.3), consistent with the preference of M-C interaction to M-M interaction. Since Ni is widely used to catalyze CNT growth, this result demonstrates that the BM structure can also make an important role in CNT synthesis.

*Table 1.* Relative strength of the M-C and M-M interactions. Carbon atom adsorption energy ($E_a$) and the cohesive energy ($E_c$) of the metal (M) are in eV, and $R$ is the ratio between them.

| M  | $E_a$ | $E_c$ | $R$  | Automatically Form BM? |
|----|-------|-------|------|------------------------|
| Pd | 7.02  | 3.70  | 1.90 | NO                     |
| Au | 4.59  | 3.05  | 1.50 | NO                     |
| Cu | 4.94  | 3.49  | 1.41 | YES                    |
| Ag | 3.51  | 2.55  | 1.38 | YES                    |
| Ni | 6.84  | 5.21  | 1.31 | YES                    |
| Co | 6.61  | 5.53  | 1.20 | NO                     |
| Ir | 7.27  | 7.30  | 1.00 | NO                     |
| Ru | 7.08  | 7.60  | 0.93 | NO                     |

Interestingly, Pd has an even larger $R$ value, but it does not automatically form a BM structure. We suggest that, in this case, the M-C interaction is too strong, which makes the adsorbed system very rigid. Therefore, the relaxation to BM structures can not happen automatically. Such a picture is confirmed by an examination on the BM structure of Pd, which is found to be a stable structure. It is 0.92 eV lower in energy than the corresponding *nnn* structure, while there is a 0.34 eV barrier between them (Figure 4). Au is an intermediate case, where its $R$ value (1.50) is only moderately larger than that of Cu, Ag, and Ni. In the optimized structure, the shared Au atom is significantly upshifted in the optimized geometry, but there is no coordination number change observed (Figure 5). Therefore, the automatic formation of BM structures requires a delicate balance between M-C and M-M interactions.

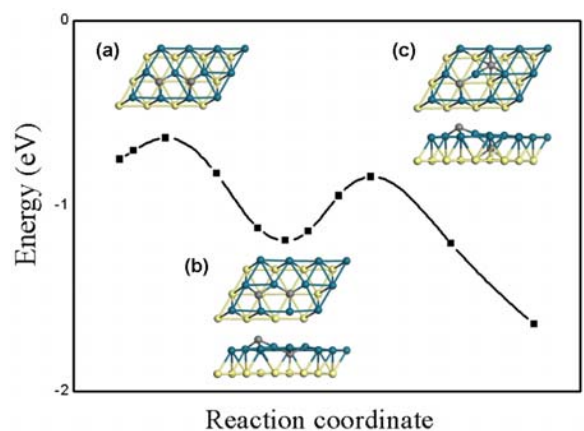

*Figure 4.* The reaction path from (a) two *nnn* carbon atoms on Pd (111) surface, passing (b) an intermediate, to (c) the BM structure.

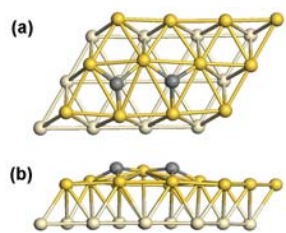

*Figure 5.* (a) Top view and (b) side view of the optimized geometry of two *nnn* carbon atoms on Au (111) surface.

To also take the temperature and C coverage into account, we perform MD simulations for carbon atoms on Cu (111), Ir (111), and Pd (111) surfaces. Initially, 1/3 monolayer of reasonably separated carbon atoms was put on the surface. Canonical ensemble simulations have been performed at 300 K for 1 ps and then at 800 K for 5 ps.

On Cu surface, carbon atoms begin to strongly diffuse even at 300 K, and the BM structure can be observed at as early as about 0.12 ps. At 800 K, the carbon coordination numbers are generally increased to four or five, and the BM structure becomes very popular on the rough surface (Figure 6a). It is interesting to note that, although there is no carbide phase existing in the bulk Cu-C phase diagram,[15] Cu and C can be well mixed on the surface at high temperatures. During the whole 6 ps simulation time, no carbon dimer is observed on the surface, which again demonstrates that the BM structure plays an important role in kinetics of graphene growth on Cu surface.

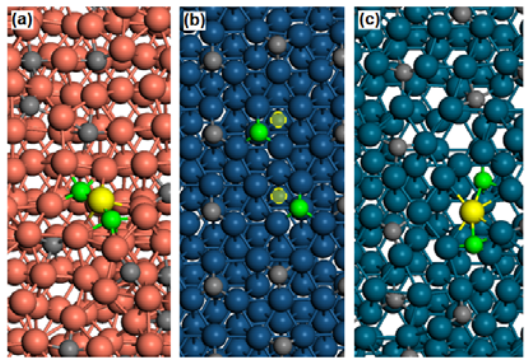

*Figure 6.* Typical snapshots in the 800 K trajectories for (a) Cu, (b) Ir, and (c) Pd surfaces. In the Cu and Pd cases, a typical BM structure is highlighted. Carbon diffusions on Ir surface are marked.

As a contrast, on Ir surface, we have not observed much structure relaxation in the whole trajectory except limited carbon diffusions (Figure 6b). This is a reflection of the relatively weak Ir-C interaction compared with Ir-Ir interaction. For Pd, in the 1 ps 300 K trajectory, all carbon atoms are trapped at their initial hollow sites. While intense surface relaxation is observed at 800 K, with some BM structures formed (Figure 6c). Therefore, MD simulations support our model very well.

The results obtained here suggest that the nucleation and growth of graphene on Cu surface are more complicated than previously expected.[5,7,8] For example, using plasma enhanced CVD or directly depositing carbon atoms to Cu surface will not lead to a very low temperature growth of graphene. If atomic carbon is immediately available, a carefully controlled growth temperature may lead to new Cu-C hybrid two dimensional structures. The universality of the BM structure also makes it possibly relevant in many surface reactions involving atomic carbon species, such as the Fischer-Tropsch synthesis. At the same time, the automatic formation of BM structure on Cu, Ag, and Ni surfaces strongly encourages a direct low temperature experimental observation. As shown in Figure 7, the simulated STM image of the BM structure on Cu surface can be easily distinguished from those of monomer and dimer. Therefore, it is readily to be identified in an STM experiment.

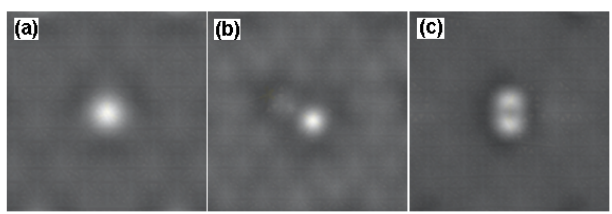

*Figure 7.* Simulated STM images of (a) a carbon adatom, (b) a BM structure, and (c) a carbon dimer on Cu (111) surface. Bias voltage is set to 0.5 V.

In summary, a new BM structural motif has been identified, which induces significant barriers during the collision of carbon species on Cu surface. This result demands an explicit consideration of carbon incorporation in future studies on graphene or CNT growth. Based on the simple homo- and hetero-interaction competition model, a general picture is obtained for a variety of surface chemistries.

**Supporting Information Available:** More computational details, energy values, and reaction paths. This material is available free of charge via the Internet at http://pubs.acs.org.

**Acknowledgment.** This work is partially supported by NSFC (20933006, 20803071, 50721091), by MOE (FANEDD-2007B23, NCET-08-0521), by MOST (2006CB922004), by CAS (KJCX2-YW-W22), and by USTC-SCC, SCCAS, and Shanghai Supercomputer Center.

**AUTHOR INFORMATION.**

Corresponding Author:

*To whom correspondence should be addressed. Email: zyli@ustc.edu.cn

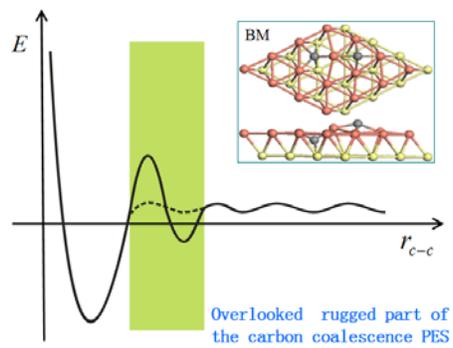

Overlooked rugged part of
the carbon coalescence PES